# Infrastructure Resilience Curves: Performance Measures and Summary Metrics

Craig Poulin [a], Michael Kane [a]*

[a] *Department of Civil and Environmental Engineering, Northeastern University, Boston, MA 02115, United States*

\* *Corresponding author. Email address:* michael.kane@northeastern.edu

## Abstract

Resilience curves are used to communicate quantitative and qualitative aspects of system behavior and resilience to stakeholders of critical infrastructure. Generally, these curves illustrate the evolution of system performance before, during, and after a disruption. As simple as these curves may appear, the literature contains underexplored nuance when defining "performance" and comparing curves with summary metrics. Through a critical review of 273 publications, this manuscript aims to define a common vocabulary for practitioners and researchers that will improve the use of resilience curves as a tool for assessing and designing resilient infrastructure. This vocabulary includes a taxonomy of resilience curve performance measures as well as a taxonomy of summary metrics. In addition, this review synthesizes a framework for examining assumptions of resilience analysis that are often implicit or unexamined in the practice and literature. From this vocabulary and framework comes recommendations including broader adoption of productivity measures; additional research on endogenous performance targets and thresholds; deliberate consideration of curve milestones when defining summary metrics; and cautionary fundamental flaws that may arise when condensing an ensemble of resilience curves into an "expected" trajectory.

# 1 Introduction

This manuscript adopts the following general definition of *infrastructure resilience*: a system's ability to withstand, respond to, and recover from disruptions [1]. This ability can be described in terms of both time and system performance [2]. A *resilience curve,* as illustrated in Fig. 1, illustrates changes over time for a selected *performance measure* within a specific scenario. A curve typically begins at a nominal level, decreases due to a disruption, then recovers (ideally back to the nominal level). *Summary metrics* are used to compare curves by quantifying key dimensions of the curves (e.g., residual performance and disruption duration in Fig. 1).

Resilience curves are applied across the critical infrastructure literature. Some implementations are qualitative or conceptual: a context for wider discussion [3]–[7], a justification for a related analysis [8]–[13], or an analysis on a specific curve element [14]–[21]. More commonly, resilience curves are implemented as the basis of quantitative analysis: historical post-disaster recovery review [22]–[32], identification of critical system properties or components [33]–[38], optimization of recovery activities [39]–[50], or comparison of system configurations [51]–[58]. Within this manuscript, *resilience analysis* refers to both modeling and empirical studies, with the former being vastly more common.

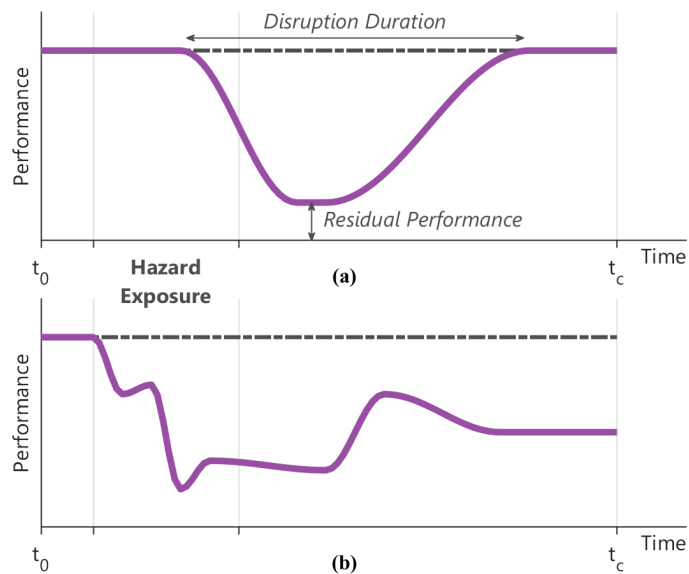

***Fig. 1.*** *Two resilience curves: (a) typical representation with two potential summary metrics, (b) a curve exemplifying non-idealized system behavior*

The resilience curve was first introduced in 2003 by Bruneau et al. [59]. Later termed the "resilience triangle" [60], the initial paradigm included instantaneous performance loss, immediate response, and a relatively linear recovery trajectory. While subsequent literature expanded the resilience triangle to a wider variety of trajectory characteristics [55], [61]–[67], this manuscript adopts a more general view of resilience curves. As in the "resilience trapezoid" paradigms [52], [53], [68], [69], system behavior may include phases of cascading failure (i.e., "absorb" [70]–[72]) and pre-recovery degraded performance (i.e., "response" [73] or "endure"). Additionally, a recovery phase may be non-increasing [33], [57], [74]–[76] and full restoration might not be possible [77], [78], as in Fig. 1(b). Illustrations within this manuscript do not include all possible variation, but the synthesis and recommendations deliberately encompass all resilience curve forms.

Formally, a resilience curve shows the evolution of a performance measure, $\mathcal{P}(t)$, which itself maps system states $x \in \mathcal{X}$ to a scalar value for all times in the scenario: $\mathcal{P}(t): \mathcal{X} \mapsto \mathbb{R}, t \in [t_0, t_c]$. A summary metric, $J$, maps an entire resilience curve to a scalar value: $J: \mathcal{P} \mapsto \mathbb{R}$. Both performance measures and summary metrics should reflect stakeholder interests and goals; analysts have many options for both. It should be noted that a measure or metric that represents stakeholder interest for one system and scenario may not accurately quantify those same interest under a different system or scenario.

Resilience curves can reflect empirical or simulated system behavior. While experimentation is possible for some computer networks [79]–[83], most empirical data comes from historical events. Examples include the 2011 Tohoku earthquake [24], [25], [27], 2010 Chilean earthquake [25], [26], 2003 Italy blackout [29]; Hurricanes Sandy and Irene [30]; and Hurricane Katrina [31], [32]. Simulated resilience curves can be generated through a variety of modeling methods. Examples include stock and flow dynamics [84], [85], network topology [49], [86], agent-based modeling [87], [88], Boolean networks [89], and discrete event simulation [90]. The details specific to each simulation technique are outside the scope of this manuscript; instead, this manuscript aims to provide generally applicable insights and recommendations.

Analysts, designers, and stakeholders should carefully select performance measures and summary metrics, since different measures or metrics can yield significantly different recommendations. Considering two disruption scenarios for a simple system, Henry and Ramirez-Marquez illustrated that different measures suggest different restoration strategies [68]. Cimellaro et al. compared "customers served" and "tank water level" as measures to quantify water system performance, highlighting that they may diverge [76]. Evaluating a traffic simulation, Nieves-Melendez and De La Garza compared threshold-based and cumulative performance-based metrics, finding significantly different results [91]. Highlighting dissimilar curves that provide equal metrics, Bruneau and Reinhorn stressed that singular metrics "should be used with care" [92]. Sharma et al. similarly warned that "a single metric cannot generally replace a curve and capture all of its characteristics" [93]. Despite such examples and caution, the literature lacks clear and general recommendations for selecting appropriate measures based on characteristics of stakeholder values and goals.

Stakeholders include "decision makers, the public, and other end users" [54]. Performance measures could be selected through expert elicitation [48], [94]. Chang and Shinozuku established objectives in consultation with peers, but acknowledged formal investigation may be warranted [54]. A RAND report describes how deliberate stakeholder engagement informs metric selection [95]. Despite these examples, resilience literature does not often address the selection of measures or metrics, perhaps because their selection is a non-trivial analytical burden. Cimellaro et al. were frank: "the authors do not want to enter the discussion of which [water quality] index is better to adopt" [76]. Additionally, potential measures vary in their subsequent level of analytical effort—either due to model formulation or data collection. Generally underexplored is the connection between types of measures, their applicability, and their consequence on analysis effort. This manuscript aims to clarify these relationships.

This manuscript is a survey of the resilience curve as an analytical tool; wider considerations of infrastructure resilience and general (non-curve) metrics are outside its scope. In this, it is unique among infrastructure resilience surveys, of which there are many. Existing surveys vary in their purpose, such as definitions, domains, methodologies, attributes, and metrics. Some do not include resilience curves [96]–[102]. Others include resilience curves as motivation or one of many conceptual considerations, but without in-depth attention [103]–[114]. Surveys devoted to metrics often summarize curve-derived metrics as one category of many [115]–[117]. In contrast, this survey does not consider non-curve metrics (e.g., network topology or qualitative attributes). Additionally, while this survey is deliberately broad, many others focus on a specific infrastructure domain [97], [99], [105], [109], [111]–[114], [117]. Each survey provides value, but there are limitations to a single publication's scope. Sun et al. provided an excellent survey of transportation measures and metrics [118], but they adopted the resilience triangle paradigm and omitted fundamental metric types defined herein. Yodo and Wang reviewed curve metrics [119], but they focused on design implications and did not address measures. Finally, many publications summarize existing work when proposing their own framework. Sharma et al. exclusively described integral-based metrics, and their proposed metric presumes instantaneous shocks and non-decreasing recovery [93]. Shen et al. established four metric categories in support of their unifying framework, but did not discuss implementation considerations [120].

In contrast to related work, this manuscript is a broad survey of the critical infrastructure resilience literature, yet with a scope limited to the synthesis and application of resilience curve measures and metrics. Across the literature, resilience curves have demonstrated utility across varied analytical goals—from communicating concepts to post-disaster assessments to simulation-based optimization. This manuscript aims to bring clarity to their adoption and implementation. *Section 2* describes the literature review methods and sources. *Section 3* defines taxonomies of performance measures and their normalization. *Section 4* defines a taxonomy of summary metrics and discusses metrics from ensembles of metric functions, measures, and scenarios. *Section 5* synthesizes best practices on the selection of measures and metrics and discusses communication and analytical advancement.

# 2 Literature Survey

This manuscript synthesizes resilience curve measures and metrics from a "systematic search and review" of critical infrastructure resilience literature. Using broad inclusion criteria and deliberate search methodology, the scope incorporates a variety of existing work. This method allows for flexible synthesis; it is more subjective than other survey approaches, such as a systematic review [121]. As a critical review, the goal is not to consolidate all previous work, but to highlight trends and opportunities.

## 2.1 Methodology

Publications were selected in a three-step process. First, candidate publications were identified through Web of Science with the query *"TOPIC: (critical infrastructure resilience) OR TOPIC: (critical infrastructure resiliency)"*. Second, results were filtered to include only those that contained resilience curve illustrations; that is, figures with performance on the vertical axis and time on the horizontal axis. This definition of a resilience curve excludes figures with events, not time, on the horizontal axis (e.g., removal or restoration of network nodes [122]–[124]); non-decreasing accumulation on the vertical axis (e.g., total economic loss over time [125]); and work solely illustrating internal state changes to maintain unillustrated nominal performance (e.g., valve position [126]). Third, during review of the subsequent collection, referenced publications were added when cited as a source for conceptual and qualitative resilience curve approaches.

## 2.2 Sources

The Web of Science search—executed in April 2020—identified 1,518 candidate publications. Of these candidates, 1,384 (91.2%) were accessible through Northeastern University licenses. Filtering provided 184 publications (13.3% of those accessible). During the review, an additional 89 publications were identified as references, bringing the total survey scope to 273 publications.

Publications spanned infrastructure types, with examples from energy [48], [127], [128], transportation [89], [111], [129]–[131], water [132]–[134], financial [74], [135], information [12], [83], [136]–[138], healthcare [66], [66], supply chains [139], [140], and coupled systems [22], [141]. Disruptions ranged in duration from minutes to hours to days, including both natural [30], [85], [142], [143] and manmade events [144], [145]. Some publications were deliberately generic for systems and/or disruptions [78], [146], [147]. Within the collection, some resilience curve illustrations were conceptual, without specified measures or metrics [70], [72], [106], [148]–[166]. Publication dates ranged from 2003 through 2020, with a steady increase after 2008. Publication types spanned 214 journal articles from 101 journals (see Table 1), 45 conference papers, 8 book sections, 5 magazine articles, and 1 report.

*Table 1* Summary of journal publications

| Publication | Count |
|---|---|
| Reliability Engineering & System Safety | 26 |
| Journal of Infrastructure Systems | 10 |
| Risk Analysis | 9 |
| Earthquake Spectra | 7 |
| Engineering Structures | 5 |
| Sustainability | 5 |
| All others (fewer than 5 articles) | 152 |

# 3 Performance Measures

This section focuses on performance measures—the vertical axis of a resilience curve. First, three categories of measures are defined: *availability*, *productivity*, and *quality*. Second, *ensemble measures*, derived from multiple candidate measures, are described. Third, performance normalization is discussed and framed with three schemes: *static*, *endogenous*, and *exogenous* normalization.

Performance measures are derived from system states, which may be numerous. Not all states of stakeholder interest are appropriate as performance measures. For example, the number of deployed generators is relevant to post-disaster operations [167], but does not reflect degraded performance. However, candidate measures will likely need to be down-selected before analysis (e.g., Pagano et al. illustrated water deficit from ten variables [57]). Section 3.2 discusses how candidate measures can be combined into an ensemble measure.

This manuscript includes performance measures, $\mathcal{P}(t)$, that are continuous (e.g., round-trip delay time [168]) or discrete (e.g., functional edges [87]). The primary requirement is that $\mathcal{P}(t)$ must lie in a metric space (i.e., a measure of distance must exist between any two values) which excludes qualitative measures [169], [170]. In this manuscript, $\mathcal{P}(t)$ indicates actual values and $p(t)$ indicates normalized values; both are considered within performance measures categorization. Note, this manuscript does not attempt to shed light on the specific formulations of performance measures for each application domain. Instead, it aims to define domain-agnostic characteristics, and provide insights into these characteristics that will facilitate outside discussions on defining performance measures for each application.

## 3.1 Categorization of Performance Measures

Infrastructure systems exist to provide service [171]. Measures can quantify system *availability*, system *productivity*, or service *quality*. Table 2 summarizes these three categories and their characteristics.

*Table 2 Performance measure categories and their typical characteristics*

|  | **Availability** | **Productivity** | **Quality** |
|---|---|---|---|
| **Description** | Capacity or aggregated functionality of the system | Quantity of service provided by the system | Character of service provided by the system |
| **Units** | Often a count; higher values desired | Often flow or rate; higher values desired | Wide variation; lower values may be desired |
| **Nominal** | Full function | All service demand met | Typical or steady-state values may provide a reference but not a bound |
| **Examples** | Customers with service [23], [24], functional cranes [89], transportation system capacity [94] | Electrical demand met [50], supply chain output [172], water demand satisfied [33] | Average vehicle speed [173], communication delay [174], water rationing level [133] |
| **Normalization** | Static reference values may be sufficient | Exogenous performance targets may be needed to represent scenario service-demand | Reference may be static or an endogenous baseline; unnormalized values may be preferred |
| **Analysis Focus** | The infrastructure system itself | Provision of infrastructure services | Utilization of infrastructure services |
| **Potential Stakeholders** | Utility operators, public works departments | Customers, public officials | Community members, operators of supported systems |
| **Scenario Applicability** | Productivity and quality are tightly coupled with availability; no anticipated changes in stakeholder goals within scenario duration | Service demand is expected to change within scenario duration; degraded conditions preclude quality concerns | Service availability and productivity are sufficient, quality is the primary concern within the scenario; comparing tradeoffs with steady-state performance |
| **Modeling Scope** | Modeling may be primarily focused on the infrastructure system | Modeling should include representation of service demand | Modeling likely includes representation of service demand; may be primarily focused on non-infrastructure considerations |

*Availability* measures describe capacity or functionality of an infrastructure system. Across literature, this is the most common category. In its simplest implementation, components are functional or non-functional, and the measure describes "number of functional components" (e.g., power buses [69], [175], cranes at a seaport [89], water system nodes [142], [176], population with power [177]). This category also applies to components with partial functionality (e.g., bridge loading capacity [18], [178]) or weighted aggregates (e.g., generation capacity [20], [179], transportation system capacity [94], [180]). Availability measures generally have a nominal value and upper bound, i.e., the assessment when all components are functional.

Varying demand for the service the infrastructure provides (i.e., service demand) may affect availability measures during a scenario. For example, demand for electricity impacts the voltage in electricity distribution systems, and as a result of low-voltage cut-offs, could affect households with service. Alternatively, many availability measures can be determined independently of service demand and its dynamics. This includes network topology measures, such as functional edges [87], path availability [181], giant component size [182], average shortest path length [143], [182], and average maximum flow capacity [35]. When the dynamics of service demand can be excluded, availability measures are generally the most straightforward category to model. Thus, availability measures may be appropriate when service demand is constant or independent of system availability. This category is best suited for stakeholders interested in the infrastructure system itself (e.g., utility operators) and not the end-use of the service provided.

*Productivity* measures describe the quantity of service provided by an infrastructure system. Service quantity is a function of supply, capacity, and demand; all three may need to be modeled. Productivity measures best represent the common definition of infrastructure, systems that "produce and distribute a continuous flow of essential goods and services" [183]. These measures are often described in terms of rate or flow (e.g., electrical load [50], [184], packet delivery ratio [145], gas supplied [185], or water demand satisfied [33]). Alternatively, productivity measures may be framed as the number of customers satisfied (e.g., computing workflows completed [186], ships berthed [187]).

Since engineers often focus on supply, productivity measures are generally expressed relative to demand (e.g., flood volume relative to rainfall [132]) or by quantifying shortfalls directly (e.g., unmet water demand [188], supply chain output [172]). In most cases, the aggregated demand, when quantified, provides an upper bound for performance at any given point in time. Productivity measures may be most appropriate when service demand is dynamic with infrastructure condition. The interests of customers and other end-use stakeholders may be best captured by productivity measures.

*Quality* measures describe the character of the service provided by an infrastructure system. Examples included hospital patient wait time [90], [189]–[191], networking throughput [79]–[82], [192], sensing coverage [193], average vehicle speed [173], and water quality index [76]. These measures generally incorporate greater context from the environment. Measures may be a proxy for attributes of interest; Cimellaro et al. adopted hospital patient wait time because "quality of care is affected by the level of crowding in the emergency room" [189]. Note, some publications labeled their vertical axis "quality" or $Q(t)$ without any relationship to this taxonomy [32], [59], [162], [194], [195].

Across the literature, a wide variety of units and presentations are used for quality measures. For some quality measures, lower values are desired (e.g., transportation travel time [196]–[199]). Authors implemented such measures with their reciprocal [196], inverted the vertical axes [197], or presented unadjusted values in contradiction to resilience curve conventions [199]. A reference value is often helpful to provide perspective to a resilience curve; however, for quality measures, this reference may be difficult to define. Reference nominal performance may take the form of typical, average [91], [173], steady-state, or planned performance [30]. Alternatively, reference target performance could reflect ideal or maximum performance, even if such performance is not expected during non-disrupted conditions. Examples include zero patient wait time [90] and communication delay [174].

Quality measures can highlight tradeoffs between steady-state and disrupted performance that may not be apparent from availability or productivity measures. Citing work by Ganin et al. [200], Linkov and Trump illustrated how steady-state traffic in Los Angles was worse than Jacksonville (as measured by commuter delay), but was more resilient when disrupted [2]. This added analytical power of quality measures, in addition to explicitly quantifying the character of service provided to end-users, could motivate their adoption within an analysis. However, these benefits need to be balanced with the complexity of not only modeling the effect of service supply and demand (as with productivity measures) but with their effect on the quality of the service. Therefore, quality measures are most appropriate when the character of provided service is the overriding consideration.

### 3.2 Multiple and Ensemble Measures

As highlighted by Bruneau et al. in their foundational work: "dimensions of community resilience…cannot be adequately measured by an single measure of performance" [59]. Every system can have multiple candidate measures (e.g., connectivity, size of giant component, and shortest path length for power, radio, and communication [201]) and subsystem measures can be calculated separately (e.g., subsystems defined by owner or spatial regions [202]). The three categories of measures defined above can provide structure when defining a set or ensemble.

Availability measures, when applied to the same subsystem, are likely to be monotonic transformations of one another. Sequential restoration of network links improves all topology-based measures, just at different rates (e.g. functioning links and maximum flow [87]). Monotonicity reduces, but does not eliminate, the impact of adopting one measure over another. In a simple network, shortest path and total functional length can each recommend different restoration strategies [68]. For non-availability measures or multiple subsystems, candidate measures may not be monotonic transformations of one another. In most disruption-then-recovery scenarios, measures will often parallel one another (e.g., functional cranes, seaport productivity, and servicing time [187]), but some may diverge (e.g., served customers and water tank reserves [76], economic sector recovery [203], [204]). Identifying, interpreting, and balancing diverging performance measures and contradictory recommendations are generally underexplored across the literature.

Multiple measures can be plotted separately [20], [80], [83], [201], [205] or together [25], [27], [133]. In general, there were three approaches to incorporating multiple measures into a resilience assessment. (1) Results of each measure can be left for open-ended interpretation by the stakeholder. (2) Summary metrics can be separately derived for each measure, and those metrics synthesized into an ensemble metric (discussed in *Section 4.7*). Or, (3) multiple measures can be consolidated into a single ***ensemble measure***.

An ensemble measure may be defined by shared or converted units. Carvalho et al. measured supply chain performance as the sum of production, material, and transportation costs [140]. Economic analyses translated sector "inoperability" to financial impact, providing a system-wide curve [204], [206]–[210]. Unit conversions may not be static. Goldbeck et al. converted transportation and energy to "value of service" with fixed parameters, but acknowledged a need for "more sophisticated" models and data [86]. Brozović et al. incorporated the price elasticity of demand for residential water [133]. To emphasize early restoration, Ulusan and Ergun implemented a "benefit function" that decreases non-linearly over the scenario [49].

Alternatively, an ensemble measure may be determined through unitless mathematical functions. In some cases, constituent measures lacked deliberate weighting. Examples were found across application domains in the literature: for water systems, the ratio of delivery capacity to shortest path length [211]; for petroleum infrastructure, the sum of consumption and storage deficits [212]; for a "power resilience index," the product of available transformer ratings, percent of undamaged substations, alternative paths, and available generators [213]. Alternatively, constituent measures may be weighted. Cimellaro et al. calculated the weighted sum of normalized flow and service area in a gas network, but asserted that weighting has little impact on recommendations [128]. In contrast, Thekdi and Santos found the value of investment alternatives varied widely with weighting across five stakeholders and five measures [94]. He and Cha implemented an "integrated network" measure as the weighted sum of topology-based measures for power, radio, and communication; they demonstrated that weighting schemes affect restoration decisions [201].

Weighting for ensemble measures should reflect stakeholder values, but weight selection and validation is underexplored. Some publications proposed weighting for customer priority or criticality, but did not recommended values nor methods [14]–[16]. Zhang et al. weighted electricity 2x greater than water [176]; Najafi et al. used a 9x factor, but acknowledged research is warranted [142]. For hospital system performance, Hassan and Mahmoud recommended the weighted sum of functionality and quality, while assuming equal weighting "for simplicity" [190]. Weighting may be established by expert elicitation [48], [202], but values may also vary between and within scenarios. Jacques et al. proposed expert elicitation to determine weighting between hospital services, but assumed equal values in their case study [214]. Ottenburger and Bai asserted that weighting for urban resilience is "highly dependent on local urban circumstances and conditions" [215]. Massaro et al. acknowledged epidemic subpopulation weighting varies with the scenario, system, costs, and stakeholder perspectives [216]. Weighting parameters may be fixed (i.e., static), driven by external factors (i.e., exogenous), or driven by internal dynamics (i.e., endogenous).

## 3.3 Performance Normalization

Performance measures are commonly normalized to facilitate comparisons across systems and scenarios. Within this manuscript, $\mathcal{P}(t)$ denotes unnormalized performance and $p(t)$ denotes normalized performance, $p = \mathcal{P}/\mathcal{R}$, where $\mathcal{R}$ is the reference, target, baseline, or nominal value that may or may not be time varying. Unnormalized values were observed in the literature for each of the three categories of measures: availability [87], [169], [177], [217], productivity [184], [185], [212], and quality [30], [79]–[82], [90], [173], [191], [193], [218]. Unnormalized values are appropriate when $\mathcal{P}$ provides key context or the nominal value is unclear or irrelevant—which is often the case for quality measures. A vast majority of resilience curves are presented in terms of $p(t)$. Such normalization can quickly communicate relative performance to stakeholders in conceptual conversations. For quantitative analysis, normalization enables comparison and optimization across scenarios and systems (e.g. comparing three electric utilities following Hurricane Sandy [219] or recovery of lifeline systems across 12 Japanese prefectures following the 2011 Tohoku Earthquake [23]). However, care should be taken when presenting only normalized values, as normalization can obfuscate important context (e.g., populations within Japanese prefectures vary by nearly an order of magnitude: over 9 million in Kanagawa to less than 1 million in Akita). This disadvantage is easily resolved by presenting both actual and normalized values.

Critically underexplored are the specific functions used to normalize $\mathcal{P}(t)$ to $p(t)$. In many cases, normalized measures are assumed or built into proposed frameworks (e.g., the resilience triangle [59], dynamic inoperability input-output model [206]). Some publications did not explicitly state their denominator (especially "fraction of customers with service" [23], [24], [26], [32], [219], [220]). Others adopted a fixed denominator without elaboration (e.g., "usually a constant" [221], "assumed no change" [222]). But analyses should not neglect the normalization reference—when an assessment is based on $p(t)$, changes in the denominator can be as impactful as changes in $\mathcal{P}(t)$. This section aims to clarify performance normalization by defining three normalization schemes: static, exogenous, and endogenous.

Across the literature, references (i.e., denominators) for performance normalization were comprised of three types, each of which is illustrated in Fig. 2. A fixed reference value, $\mathcal{R}_0$, does not change within the scenario; this provides ***static normalization***, $p(t) = \mathcal{P}(t)/\mathcal{R}_0$. A time-varying baseline, $\widehat{\mathcal{R}}(t)$, changes in time but does not respond to the scenario; this provides ***exogenous normalization***, $p(t) = \mathcal{P}(t)/\widehat{\mathcal{R}}(t)$. A time-varying performance target, $\mathcal{R}(t)$, changes in time and responds to scenario and system dynamics; this provides ***endogenous normalization***, $p(t) = \mathcal{P}(t)/\mathcal{R}(t)$.

**Static normalization**, illustrated in Fig. 2(a)-(b), incorporates a fixed reference value, $\mathcal{R}_0$. Availability measures were normalized almost exclusively with this scheme. For availability measures, $\mathcal{R}_0$ was the nominal value (and generally the upper bound); static normalization emphasized restoration to full function, regardless of the measure [38], [40], [44], [53], [55], [76], [87], [133], [143], [176], [182], [201], [216], [219], [223]–[225]. While the $\mathcal{R}_0$ reference may change over time (e.g., population growth, infrastructure expansion, and system modification), the time scale of such changes is typically outside analyzed resilience scenarios.

Static normalization was also applied to quality measures, but was less straightforward. When $\mathcal{R}_0$ describes typical or expected performance, $p(t)$ may exceed one (e.g., normalizing to the speed limit [91]). Alternatively, when $\mathcal{R}_0$ describes an extreme upper bound, steady-state $p(t)$ may be less than one (e.g., normalizing to zero for patient wait time [90], [190] or communication delay [174]). When lower values of performance are desired (e.g., wait time), the normalized value may be inverted $p(t) = \mathcal{R}_0/\mathcal{P}(t)$ [198] or the vertical axis may just be graphically flipped [226] to maintain the intuitive standard "up is good and down is bad". In cases where the system should not deviate up nor down from nominal, (e.g., electrical bus voltage [227]) the absolute deviation or square could be minimized. None of these considerations are insurmountable, but they make normalization of quality measures distinct from that of availability measures.

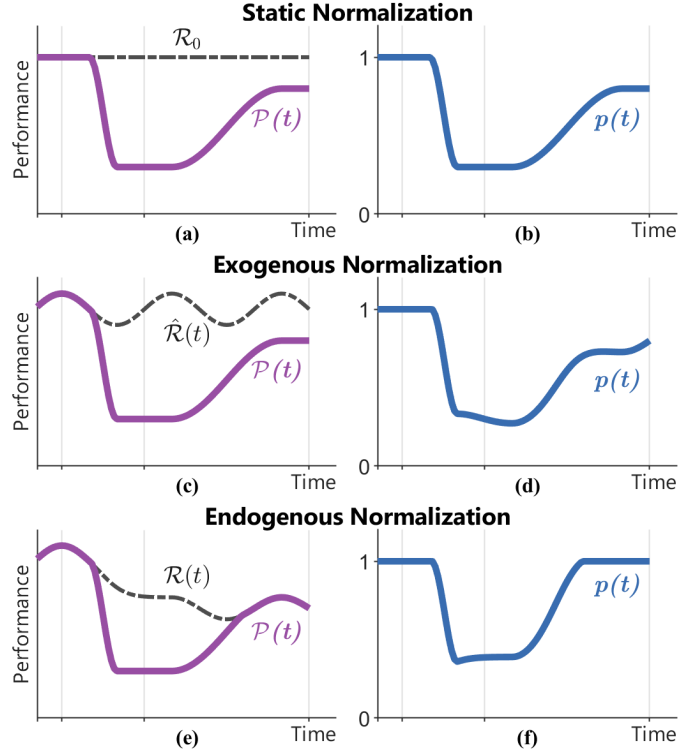

*Fig. 2. Comparison of normalization schemes used to translate actual performance measures, $\mathcal{P}(t)$, to normalized performance, $p(t)$. Static normalization (a-b) incorporates a fixed reference value, $\mathcal{R}_0$. Exogenous normalization (c-d) incorporates a time-varying baseline, $\hat{\mathcal{R}}(t)$, that does not adapt to the scenario. Endogenous normalization (e-f) incorporates a performance target, $\mathcal{R}(t)$, that responds to scenario dynamics. While $\mathcal{R}(t)$ is shown to approach $\mathcal{P}(t)$ in (e), this is not a requirement or expectation of performance targets; modeling $\mathcal{R}(t)$ may be as intensive as modeling $\mathcal{P}(t)$.*

Across the literature, static normalization was applied to productivity measures, but its applicability is less clear. Productivity measures describe a rate or flow of service, so a fixed $\mathcal{R}_0$ implies service demand does not vary in time. Static normalization of productivity measures included packet delivery ratio [145], economic inoperability [203], transported supplies [49], water delivered [205], and electricity provided [143], [228]. Many real-world systems vary service demand over time, under both steady-state and disrupted conditions; flattening such variation affects $p(t)$ and may impact analytical recommendations. In contrast, time-varying references may be either exogenous or endogenous.

**Exogenous normalization**, illustrated in Fig. 2(c)-(d), incorporates a time-varying baseline $\hat{\mathcal{R}}(t)$ that does not adapt to the scenario (i.e., the hazard nor the corresponding system response) "without the effects of hazard" [187] or "under no disruption" [51]. No availability measures implemented exogenous normalization; the only quality measure was profit, baselined to cyclical customer demand [21]. Static normalization can be considered a special case of exogenous normalization: $\hat{\mathcal{R}}(t) = \mathcal{R}_0 \; \forall \; t \in [t_0, t_c]$. Rose made this connection explicit by abstracting economic growth "for ease of exposition and without loss of generality" [74]. Other authors defined productivity normalization with a time-varying reference but implemented a fixed value in their case study [34], [50], [56], [229], [230]. Despite this connection, the manuscript treats static and exogenous normalization as distinct schemes.

Productivity measures widely implemented exogenous normalization. A time-varying baseline can address natural variations in service demand (e.g., hourly, daily). Examples included typical or historical traffic levels [30], [86], [231], [232], electrical demand [58], [86], [176], and water consumption [75], [176]. Cimellaro et al. illustrated how timing of system failures affects summary metrics due to hourly changes in water demand [76]. However, exogenous normalization does not incorporate dynamics between the scenario and service demand. Often, performance is assumed not to exceed the reference; however, this assumption may be invalidated by exogenous normalization. For example, Shafieezadeh and Ivey Burden normalized seaport offloading to an exogenous baseline; they observed that $p(t)$ exceeded one as delayed processes were eventually accomplished during a cyclical lull in demand [187]. Such peculiarities need not be a problem, but should be anticipated when adopting exogenous normalization for productivity measures.

**Endogenous normalization**, illustrated in Fig. 2(e)-(f), is the most general scheme. It incorporates a time-varying performance target, $\mathcal{R}(t)$ that is affected by the hazard and/or the corresponding response of the system or coupled-systems. If $\mathcal{R}(t)$ depends only on the hazard or the open-loop response of a coupled system to the hazard, $\mathcal{R}(t)$ may be decoupled from main-system dynamics, and may even result in a constant value for a given hazard. Generally, implementing $\mathcal{R}(t)$ expands the scope of an analysis. For example, warning events [78] could serve to adjust transportation demand before hurricane landfall [233], [234]. Simulating $\mathcal{R}(t)$ almost certainly requires more effort than $\hat{\mathcal{R}}(t)$ or $\mathcal{R}_0$ as its underlying mechanisms may differ from those of $\mathcal{P}(t)$. Despite this burden, examples for all three performance measure categories highlight the applicability of endogenous normalization.

Availability measures adopted endogenous normalization when full recovery was not possible or feasible. Examples included discounting "permanently lost customers" in historical utility restoration [26] and isolating deaths "due to health care capacity" following a seismic event [92]. Such targets reflect shifts in stakeholder goals, which may only be clear in hindsight. The distinction can be significant—three years after Hurricane Katrina, some regions had 50% of their pre-hurricane electric customers [235]. In a related technique, some authors normalized to the initial performance drop (e.g., to compare power and communication recovery in the same earthquake [25]). More commonly, this approach is presented as the *restoration ratio* metric described in *Section 4.8*.

Productivity measures are ideal candidates for endogenous normalization. Just as service demand varies in time, service demand may change within a disruption scenario. Service demand could respond to the initiating hazard, such as emergency services after an earthquake [49], [194] or traffic during floods [130]. But even limited dynamics may be challenging to model; some publications highlighted difficulties and implemented a fixed $\mathcal{R}_0$ [37], [196], [199]. Further consideration of $\mathcal{R}(t)$ introduces yet more potential dynamics, which each increase modeling complexity. For example, Brozović et al. incorporated customers' price elasticity to alter their demand in response to water shortages [133]. These dynamics may be relevant to system understanding—Pagano et al. observed that evacuations reduce water demand, increasing short-term performance but stagnating long-term recovery [57]. Additionally, $\mathcal{R}(t)$ introduces demand-side opportunities to improve resilience; for example, electrical load management can shed load to avoid overloading transmission infrastructure [84]. Despite these examples, general recommendations for use of endogenously normalized performance measures remain underexplored across the literature.

Endogenous normalization for quality measures may be applied when standards of service vary during distinct phases of the scenario. Although not quantified, Cutts et al. proposed a "resilience prism" that distinguishes post-disaster community goals over days, weeks, and years [236]; specific timing of transitions between survival, safety, and belonging could be driven by emergent system behavior. Additionally, quality measures are likely influenced by dynamic service demand, even if not captured within $\mathcal{R}(t)$. For example, calculating travel times requires origin-destination pairs (i.e., transportation demand) [198]—due to modeling challenges, Chang et al. ruled out travel delay as a performance measure [18]. As a general assertion: if service demand dynamics challenge normalization of productivity measures, then quality measures are similarly constrained.

These normalization schemes provide insight into "adaption" phases of resilience. Within the literature, some resilience curves were illustrated with a post-recovery performance above the baseline [70], [72], [114], [237], [238]. Such a period could reflect rebuilding to higher standard or incorporating new information (e.g. post-hurricane investment [56]). For quantitative analyses, such adaption can only be related to $\mathcal{R}_0$ or $\hat{\mathcal{R}}(t)$. Exceeding $\mathcal{R}(t)$ would be impossible or undesired for most productivity measures. For quality measures, $\mathcal{R}_0$ might reflect typical or expected performance, in which case, system adaption would result in a new $\mathcal{R}_0$ in the next scenario. This is not criticism of "adapt" phases but commentary on quantitative limitations, especially if expressed within a summary metric.

# 4 Summary Metrics

Summary metrics map a resilience curve to a scalar value: $J\big(\mathcal{P}(t)\big)$ or $J\big(p(t)\big)$. Metrics enable comparison of system behavior across scenarios and configurations. To support the selection and implementation of metrics, this section defines six categories, provides examples, and synthesizes their best practices. The taxonomy is summarized in Table 3 with key examples shown in Fig. 3. Within the survey, there was no consensus on the "best" metric; instead, metrics were commonly linked to desired attributes. Only one publication specifically compared metrics: Nieves-Melendez and De La Garza evaluated a traffic system with threshold- and integral-based metrics, each providing a different recommendation [91].

*Table 3* Summary metric categories

| | Examples | Formulation Considerations |
|---|---|---|
| **Magnitude** (§4.2) | Residual performance, depth of impact, restored performance | - Units of performance<br>- Milestone or criterion for which performance is assessed<br>- Reference value, if presented as a ratio |
| **Duration** (§4.3) | Disruption duration, resistive duration | - Units of time<br>- Milestones or criteria that establish endpoints |
| **Integral** (§4.4) | Cumulative impact, cumulative performance* | - Units of performance × time<br>- Global control time<br>- Performance target, if normalized |
| **Rate** (§4.5) | Failure rate, recovery rate | - Units of performance over time<br>- Milestones or criteria that establish endpoints<br>- Method to address non-linear segments |
| **Threshold** (§4.6) | Threshold adherence*, residual capacity | - Units either binary or modification of another metric |
| **Ensemble** (§4.7) | Weighted sum of metrics, weighted product of metrics | - Generally unitless, expressed as an index<br>- Weighting between constituent metrics<br>- Conversion of dissimilar units |

*\* for clarity, common "resilience" terminology is not recommended*

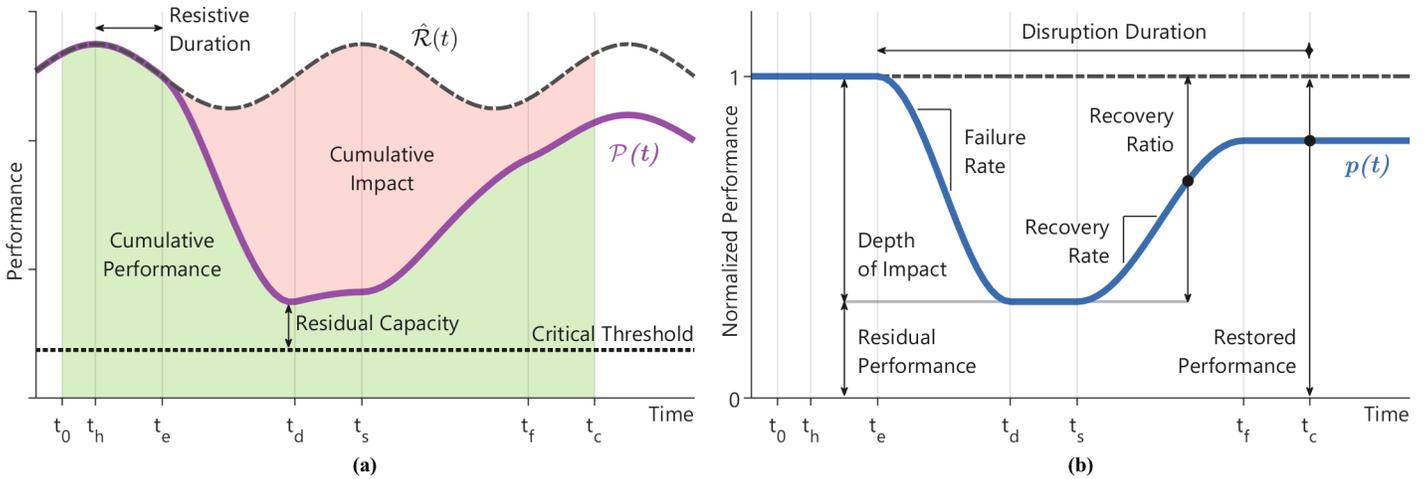

**Fig. 3.** Illustrative summary metrics for a resilience curve in terms of (a) actual units, $\mathcal{P}(t)$, and (b) normalized units, $p(t)$. Examples are not definitive; each metric can vary in its definition. In this illustration, the system does not fully recover within the control interval, so disruptive duration may be undefined. The curve does not fall below the critical threshold, so "threshold adherence" is successfully met.

## 4.1 Notation, Curve Milestones, and Control Times

Metrics may be derived with respect to actual performance, $\mathcal{P}(t)$, as in Fig. 3(a), or normalized performance, $p(t)$, as in Fig. 3(b). Metrics may also incorporate the reference value ($\mathcal{R}_0$, $\hat{\mathcal{R}}(t)$, $\mathcal{R}(t)$ or 1 if normalized). Many publications implicitly adopted static normalization and defined metrics without distinguishing $p(t)$ from $\mathcal{P}(t)$. Such assumptions may disrupt extension to exogenous baselines and endogenous performance targets. In contrast, other publications specifically defined metrics in terms of $\mathcal{R}(t)$, even if implemented as static throughout the publication [21], [86], [194], [220]. Because normalized measures are most common, this section describes metrics in terms of $p(t)$.

Summary metrics are assessed over or within a scenario's control time interval, $[t_0, t_c]$. Additionally, metrics were commonly defined in terms of curve milestones, of which five were common in the literature. As illustrated in Fig. 3, each milestone delineates a transition between phases:

- exposure to a hazard, $t_h$, transitioning from *prepare* to *resist*;
- initial system disruption, $t_e$, transitioning to *absorb*;
- end of cascading failures, $t_d$, transitioning to *endure*;
- the beginning of system recovery, $t_s$, transitioning to *recovery*;
- and the completion of system recovery, $t_f$.

Specific terminology and definitions varied across the literature, but these milestones reflect common reference points (e.g., hazard exposure [58], [238]–[240] and the "resilience trapezoid" [52], [53], [68], [69]). However, milestones are not necessarily essential, comprehensive, or unambiguous. Some curves lack milestones (e.g., the "resilience triangle" [59], [60]) or have additional milestones (e.g., partial recovery [241], [242]). Some curve trajectories do not provide clear milestones (e.g., temporary performance increases [57], [74], [75], Fig. 1(b)). Resilience analysis must clearly define the milestones used to define metrics, and those definitions should be validated over the range of considered trajectories.

The control interval bounds the period of interest. Many publications established their control interval from curve milestones—e.g. $t_e$ to $t_f$ [59], [92], [143], [194], [215], [241], [243]–[246] or $t_s$ to $t_f$ [14]–[16], [47], [55], [196]. This approach is appropriate in some cases, such as recovery sequencing, but it can provide unclear metrics when comparing dissimilar curves. Some authors established $t_0$ as the moment of disruption, $t_e$ [24], [142], [205]. This is reasonable when all curves degrade at the same time, but it risks conflating hazard exposure with system disruption, undervaluing attributes of resistance [69], detectability [238] and prediction [146]. This is avoided when $t_0$ precedes the disruption [76], [187], [216], [247], [248] (i.e., providing *prepare* and/or *resist* phases [21], [71]).

The terminal control time, $t_c$, seeks to provide a "suitability long" [62] duration. Four methods for establishing $t_c$ were common in the literature. Method one: $t_c$ is the **expected or mean recovery** time [142], [198]. This approach truncates curves that extend beyond the expectation. Method two: $t_c$ is the **maximum recovery** time. This could be determined *post hoc* (e.g., from a set of Monte Carlo runs [67]) or with knowledge of underlying dynamics (e.g., "maximum extinction time" in an epidemiological model [216]), but this approach would be undefined for systems that do not fully recover. Method three: $t_c$ reflects **scenario-specific stakeholder considerations**. Examples included a region's fresh water reserves [211] or its emergency planning standard [76]. This approach will also truncate curves, but it may assist in communicating results. Method four: $t_c$ is the **lifecycle for probabilistic hazards**. This approach considers not just a single scenario, but the generation of hazards over time (e.g., 1 year [31], [69], 30 years [189], 100 years [222], or everywhere in between [56], [220], [230]). Such analyses may require additional considerations, such as discount rates and post-recovery adaptation [56], but each disruption scenario could be evaluated using summary metrics like those described in this section.

## 4.2 Magnitude-based Metrics

Magnitude-based metrics quantify performance at a specific milestone or point in time. These include *residual performance*, *depth of impact*, and *restored performance*, each shown in Fig. 3(b). Magnitude-based metrics can be described with either actual or normalized performance units.

**Residual performance** metrics describe system performance following the disruption, generally after cascading failures. In Fig. 3, this is indicated by $J = p(t_d)$. Labels included: *avoided performance drop* [248], *minimum performance* [89], [219], *residual capacity* [249], *residual functionality* [120], [189], [250]–[254], *residual performance* [222], [255], *resistance index* [256], *robustness* [58], [63], [64], [71], [155], [184], [195], [241], [257] and *static resilience* [74]. Related forms included performance relative to a critical threshold [248], [249] and the average performance during the period of disruption [248].

**Depth of impact** metrics are the complement of residual performance: $J = 1 - p(t_d)$. Labels included: *Λ-metric* [52], [53], *amplitude* [224], [258], *consequence* [211], *depth of failure propagation* [77], *drop of functionality* [189], *initial loss* [63], [64], *lost functionality* [250], *maximum impact level* [69], *maximum incurred performance loss* [219], *peak severity* [226], *performance attenuation* [259], and *risk* [71]. Residual performance and depth of impact were associated with absorptive capacity [127], [219], survivability [13], [260], vulnerability [261], and, most often, robustness [32], [67], [86], [120], [210], [225], [253], [255], [262].

Residual performance metrics require a clear definition of the milestone of interest, especially in cases like Fig. 1(b). Potential stakeholder interests may include post-hazard performance, post-cascading failure performance, or minimum performance. Dorbritz differentiated hazard impact and subsequent system degradation, distinguishing robustness from resourcefulness and internal stability [182]. Residual performance may be defined as the minimum level of performance within the control interval [86]: $J = \min p(t), t \in [t_0, t_c]$. Candidate milestones may not align (e.g., stalled recoveries [57], [74], [75]). Additionally, with endogenous performance targets (e.g., post-earthquake hospital demand), the minimum $p(t)$ could be decoupled from physical degradation, in which case the metric no longer reflects robustness attributes.

**Restored performance** metrics quantify a system's performance after recovery efforts are complete. Labels included: *degree of return* [219], *level of recovery* [263], *magnitude of loss* [264], *recovery ability* [58], [241], and *recovery degree* [211]. Additional publications included this concept without defining a metric [259], [265], [266]. Metrics were defined relative to the initial performance level: $J = p(t_f) - p(t_e)$ [219], [263]; $J = p(t_f)/p(t_e)$ [127], [211]; or $J = \left(p(t_f) - p(t_d)\right)/\left(p(t_e) - p(t_d)\right)$ [58], [241]. Restored performance metrics may be appropriate for partial recovery (e.g., "permanent outages" after Hurricane Katrina [32]) or post-recovery improvements [58], [127], [241], the calculation of which depends the reference and normalization scheme (discussed in *Section 3.3)*.

## 4.3 Duration-based Metrics

Duration-based metrics quantify time between milestones. Some authors considered **disruptive duration** (Fig. 3(b)) over the entire period of degraded performance: $J = t_f - t_e$. Labels included: *failure duration* [226], *interruption duration* [184], *rapidity* [203], [257], *recovery time* [211], [263], [264], *restoration time* [30], *time to recovery* [219], and *total length of the disruptions* [86]. Others considered the recovery phase, starting from the period of lowest performance: $J = t_f - t_d$. Labels included *rapidity* [225], [267], [268], *recovery time* [63], [64], [189], [193], [259], and *time to full repair* [87]. Others defined metrics for $[t_e, t_d]$, the absorb phase [193], [259], or $[t_d, t_s]$, the endure phase [52], [53]. Across these forms, duration-based metrics were associated with attributes of *adaptive capacity* [261], *rapidity* [86], [184], [210], [225], [262], [267], *rapidness* [142], [211], and *recoverability* [13], [260].

To compare dissimilar curves, duration-based metrics must clearly specify their starting and ending milestones. For metrics focused on disruption duration, starting milestones were clear and consistent: $t_e$ [30], [86], [184], [203], [211], [219], [226], [263], [264]. As emphasized by Erol et al., this milestone should be distinct from hazard exposure, $t_h$ [263]. For metrics focused on recovery duration, milestones were often ambiguous. Some authors conflated the end of cascading failures with the onset of recovery actions, i.e. $t_d = t_s$ [87], [193], [259]. Others described metrics with the resilience triangle paradigm, i.e. $t_e = t_d = t_s$ [63], [64], [189], [225], [267], [268]. A resilience curve's recovery time is undefined if the system never fully recovers, such as in Fig. 3(b). This can be addressed by establishing the duration from specified thresholds [209], [209], [263], such as restoration to 95% of daily transit ridership [30].

For most duration-based metrics, lower values are preferred: shorter disruptions and faster recoveries. In contrast, some forms desire higher values, such as *speed recovery factor* as the ratio of "slack time" to recovery duration [127] or *resilience* as the average percent of "uptime" for electrical loads [269]. While generally underexplored, higher values are also desired for the duration between hazard exposure and system disruption, $[t_h, t_e]$. Labeled **resistive duration** in Fig. 3(a), this metric has been associated with *absorption* [73] and *resistance* [77], [269].

## 4.4 Integral-based Metrics

Integral-based metrics incorporate both time and performance. There are two common forms, labeled in Fig. 3(a). **Cumulative impact** is the integrated difference between performance and its reference: $J = \int_{t_0}^{t_c}(\mathcal{P}(t) - \mathcal{R}(t))$. **Cumulative performance** is its complement, generally normalized to the reference: $J = \int_{t_0}^{t_c} \mathcal{P}(t) / \int_{t_0}^{t_c} \mathcal{R}(t) = \int_{t_0}^{t_c} p(t)$. Although described here with the general $\mathcal{R}(t)$, these metrics were often implemented with fixed $\mathcal{R}_0$ values. Often perceived as a holistic quantification of system resilience, this category provided the survey's most frequently adopted metrics, spanning conceptual [242], [270], [271], empirical [24], [30], [235], [247] and simulation [21], [50] applications in system assessment [33], [198] and recovery sequencing [50], [272].

**Cumulative impact** was proposed in Bruneau et al.'s foundation work as *loss of resilience* [59]. Subsequent terminology includes: *area* [52], [53], *conditional vulnerability* [253], *functional service loss* [243], *impact area* [69], [248], *loss indicator* [224], *loss of resilience* [92], [244], *performance loss* [58], [119], [197], [221], [241], [259], *residual loss* [206], [207], *resilience loss* [86], [245], [267], *service loss* [37], *severity* [226], and *systemic impact* [43], [188], [194]. Lower values indicate better system resilience.

**Cumulative performance** was commonly denoted *R* and termed *resilience* [32], [45], [47], [55], [62]–[64], [128], [187], [190], [196], [205], [215], [228], [247], [270], [273], [274], with variations including *annual resilience* [69], *dynamic resilience* [221], [275], *level of resilience* [143], *resilience index* [22]–[24], [198], [226], [276], and *system resilience* [56], [220], [230], [277]. Despite this consensus, this manuscript adopts the term *cumulative performance* for three reasons. One: "resilience" has been used for other metrics [91], [193], [243], [269], [278] and factors not related to resilience curves [279]. Two: rather than associate a broad term with a specific implementation, *cumulative performance* describes a category of metrics with varied forms (e.g., multiple disruptions [31], weighting by scenario probability [65], limited to recovery [15], [16], subtracting costs [14], [15]). Three: summary metrics inform interpretation of a resilience curve, but they do not fully describe the system. This manuscript also does not label the curve's vertical axis "resilience," as in varied alternatives [8], [52], [53], [280]–[285]. Resilience cannot be adequately described as either a single point in time or a singular summary metric—labeling any metric as "resilience" risks constraining an analysis.

Over the control duration $T_c = t_c - t_0$, integral-based metrics may be unnormalized, $\int_{t_0}^{t_c} \mathcal{P}(t)$, or normalized to time, $(1/T_c) \int_{t_0}^{t_c} \mathcal{P}(t)$, performance, $\int_{t_0}^{t_c} p(t)$, or both, $(1/T_c) \int_{t_0}^{t_c} p(t)$. Stakeholders may prefer the clarity of unnormalized metrics (e.g., *disruption days* [224], *lost service days* [30]). For power systems, unnormalized values quantify energy directly [14]–[16], [58]. Normalizing only one dimension is uncommon. Normalized time with unnormalized performance provides the average in actual units [47], [58]. Normalized performance with unnormalized time provides odds units like fractional hours [86], even if not explicitly stated. The most common approach was to normalize both performance and time. This provides a unitless metric with $J \in [0,1]$ as the typical range [71], [129], [196], [274].

Definitions most often considered a global control duration: $T_c = t_c - t_0$. However, a large control interval may obscure nuance: if a system recovers, normalized cumulative performance approaches one as the control duration approaches infinity [187]. For clarity, Bao et al. presented the control duration as the metric's independent variable, i.e. $J(T_c)$ [21]. Alternatively, some publications defined integral-based metrics over curve milestones (e.g., $t_e$ and $t_f$) [59], [92], [143], [194], [215], [241], [243]–[246]. For some analyses, such as recovery sequencing, this may be acceptable. But, for curves with different milestones, this approach may confuse comparisons; a fixed interval may be preferred (as discussed in *Section 4.1*).

In the common, straightforward implementation of integral metrics, all units (i.e., performance × time) are assumed to be equally valuable. However, as Green highlighted, this is "open for debate" [242]. Publications explored potential differences through stakeholder weighting of milestones [51], nonlinear relationships between value and disruption duration [37], [286], an exponentially decaying "benefit function" [49], and separate pre- and post-disruption evaluations [189]. These limited examples illustrate opportunities to extend integral-based metrics to better reflect stakeholder value in across a scenario.

## 4.5 Rate-based Metrics

Rate-based metrics quantify how system performance changes over time. The resilience curve's derivative, broadly, has been labeled *agility* [287], [288] or *local resilience* [289], [290]. Most commonly, rate-based metrics focus on the failure or recovery phases.

**Failure rate**, shown in Fig. 3(b), was associated with resistance [291] and adaptive capability during "graceful degradation" [219]. Labels included: *Φ-metric* [52], [53], *disruption speed* [269], *rapidity* [58], [241], *rate of departure* [249], *rate of performance loss* [219], *reduction rate* [173], and *robustness* [259]. Metrics were calculated using the curve's derivative [259], [269], linear approximation [249], and piecewise linear approximation [58], [241]. Based strictly on the curve's negative derivative, higher values—with a lower magnitude—are desired. Some metrics were presented in this way [52], [53], [259], [269], while others used the absolute value, in which lower values are desired [58], [173], [249]. This distinction is relevant if the metric is used in an ensemble.

**Recovery rate**, shown in Fig. 3(b), was associated with restorative capability [219]. Labels included: *Π-metric* [52], [53], *rapidity* [32], [58], [67], [189], [241], [252], [255], *recovery efficiency* [225], *recovery rate* [224], [258], *recovery speed* [120], *restoration rate* [219], and *restoration speed* [269]. Like failure rate metrics, recovery rates were evaluated with the derivative of the curve [120], [269], linear approximation [52], [53], [255], and piecewise linear approximation [58], [241], plus two novel extensions: arctan( ) to the linear approximation, providing $J \in [0,1]$ [252]; and fitting recovery to an exponential decay curve, then quantifying *rapidity* as the best-fit decay constant [32]. For recovery rate metrics, higher values are preferred (i.e., steeper recovery).

Rate-based metrics require clearly specified milestones. Failure rate metrics were consistently applied over $[t_e, t_d]$. Recovery rate metrics were specified to span either the recovery phase $[t_s, t_f]$ [52], [53] or the combined endure and recovery phases $[t_d, t_f]$ [252]. Rate-base metrics were often defined and illustrated with curves that begin recovery immediately, i.e. $t_s = t_d$ [58], [67], [189], [219], [225], [241], [258], [269]. For curves with less clearly defined milestones, as in Fig. 1(b), rate-based metrics may be challenging to implement consistently.

## 4.6 Threshold-based Metrics

Thresholds describe non-linear transitions in quantitative interpretation. Thresholds could be defined in either the performance or time domains of the curve: critical performance thresholds, below which performance is unacceptable [54], [63], [78], [91], [94], [103], [105], [114], [120], [135], [226], [248], [249], [268], [282], [283], [286]–[288], [292]–[295], and recovery time thresholds, within which the system should achieve specific objectives [54], [63], [65], [78], [91], [103], [120], [127], [135], [185], [271], [286], [295].

There were two categories of threshold-based metrics. *Threshold adherence* metrics provided categorical assessment of the system. Often termed "resilience," these were defined as maintaining performance above the critical threshold [282], [283], [287], [288], [293], [296], [297], recovering within the time threshold [185], [228], or remaining within both thresholds [286]. *Threshold modified* metrics adjusted forms from another category, such as overriding the calculation with zero if below the performance threshold [94], [298]. Shown in Fig. 3(a), Pflanz and Levis defined *residual capacity* as the difference between residual performance and the critical threshold [249]. Wen et al. incorporated a recovery time threshold in translating cumulative impact to *degree of resilience* [286]. Reiger defined *brittleness* as a modification of cumulative impact: the integral below a critical performance threshold [287], [288]. Thresholds indicate a transition in stakeholder value—they may also indicate a discontinuity for the application of summary metrics.

## 4.7 Ensemble Summary Metrics

Many publications sought a single value to consolidate multiple metrics for a single curve, blend multiple performance measures for the same scenario, or capture possibilities of system behavior across scenarios. Each of these cases provides an ensemble summary metric category: metric ensembles, measure ensembles, and scenario ensembles. These ensemble metrics yield a single value for optimization or succinct communication of results; however, details of each may be easily obscured or misinterpreted.

### 4.7.1 Ensembles of Summary Metric Functions

A single curve can be described with many metrics, each representing desired attributes (i.e., $J_1(p(t))$, $J_2(p(t))$, ...). Metrics may be compared in a table [58], [220] or illustrated with contour plots [63], [64], [210]. Alternatively, a **metric ensemble metric** aims to combine distinct metrics derived from the same curve: $J = f(J_1(p(t)), J_2(p(t)), ...)$. Constituent metrics generally have different units, so metric ensembles were often (for all intents) unitless. The one exception was conversion to financial units, such as translating cumulative impact and recovery time through contract provisions [20]. This also enables the incorporation of non-curve considerations (e.g. weighted sum of cumulative impact and recovery costs [43], [172], [188], [194]).

For metric ensembles, publications varied in their constituent metrics and combination methods. Cheng et al. summed cumulative performance, residual performance, restored performance, and translations of failure and recovery durations [193]. Najafi et al. summed residual performance, cumulative performance, and recovery time [142]. Francis and Bekera defined *resilience factor* as the product of residual performance, restored performance, and their duration-based *speed factor* [127]. Nan and Sansavini multiplied residual performance, recovery rate, restored performance, and the reciprocals of failure rate and cumulative impact [58], [241]. Cai et al. proposed a more complex *resilience metric* with the product of steady-state, residual, and recovery performance and natural logarithm of disruption time and recovery duration [299], [300]. Across these examples, weighting and form validation was generally underexplored. For example, while sums and products both provide common directionality between constituent metrics and their ensembles, their behavior is different at extremes. For products, any element can drive the metric to zero. This behavior may or may not be desired for a specific resilience analysis—determination lies with the stakeholders and their goals.

### 4.7.2 Ensembles of Performance Measures

For a specific system within a specific scenario, stakeholders may be interested in multiple performance measures. *Section 3.2* described how candidate measures could be consolidated into a single ensemble measure. Alternatively, each candidate measure could provide a distinct curve, each of which is summarized by the same metric. These metrics can be consolidated into a **measure ensemble metric**: $J = f(J(p_1(t)), J(p_2(t)), ...)$.

Within the literature, every instance evaluated constituent curves with an integral-based metric. With common units, measure ensembles can be a straightforward sum (e.g., "disruption days" [224], financial units [29], [133], [221], [301]). When measures have dissimilar units, normalized metrics provide a basis for combination. Examples included: geometric mean of measures' cumulative performance [275]; combination with an assumption of independence [255]; and the unweighted product of measures' cumulative performance [76]. Weighting may be necessary to reflect stakeholder preferences. Zou and Chen proposed three weighting schemes when analyzing interdependent transportation and electric systems, leaving the choice to the "view and judgement of decision-makers" [196]. Reed et al. envisioned a function that "reflects [subsystem] interdependence and connectivity" [32]. Moslehi and Reddy applied time-varying weighting for heating, cooling, and power systems reflecting the season and time of day [243]. Weighting schemes may be estimated from historical data, such as applying time-series cross-correlations functions across sectors [23], [24], [26]. Together, these approaches outline underexplored opportunities for ensemble metrics across performance measures.

### 4.7.3 Ensembles of Scenarios

Finally, resilience analyses are often interested in a system's possible responses across a suite of scenarios. Broadly, "scenarios" can reflect differences in system configuration [52]–[55], [58], [127], hazard exposure [57],

24 / 50

[69], [89], [90], [143], [191], or emergent behavior for a given configuration, environment, and hazard exposure (e.g. in Monte Carlo simulation) [86], [87], [90], [196], [252]. Across a set of scenarios, illustrating each resilience curve is not always practical. For example, Barker and Santos illustrated distinct resilience curves for two inventory configurations, but only presented summary metrics in considering five strategies [204], [208].

A **scenario ensemble** provides the summary metrics derived from each enumerated scenario or simulation run: $\{J(p(t)|s_1), J(p(t)|s_2), ...\}$. Summary statistics enable quick analysis from such a set. Cimellaro et al. described 12 water disruption scenarios with the minimum and maximum recovery times from 5,000 simulation runs [76]. From 15,000 samples over six hazard intensity levels, Landegren et al. plotted the median, mean, and 5/95 percentile values of cumulative impact [267]. Multiple authors proposed extending binary threshold metrics, quantifying the probability of remaining within both thresholds [54], [91], [120].

Some authors illustrated the scenario ensemble through a histogram or distribution function. Examples included: recovery time and cumulative impact across 500 runs [89]; cumulative performance across 200 runs for each repair crew size option [196]; cumulative impact across multiple strategies and scenarios of 1000 runs [87]; residual performance across 10,000 samples with and without earthquake aftershocks [129]; cumulative performance and recovery rate from 100,000 runs [252]. Despite these examples, this approach is infrequently adopted—it was more common for metrics to be presented as a single value.

Considering emergent behavior for a specific configuration and hazard exposure, many authors produced an "expected trajectory" from the mean performance value at each timestep. Such an "expected trajectory" does not necessarily correspond to any trajectory within the ensemble. When presented alone, such illustrations were occasionally unclear—it was not obvious the resilience curve did not represent a specific trajectory [44], [55], [187], [191], [196]. Illustrations with constituent curves, error bars, or a confidence interval were much more clear [47], [193], [200], [216], [217], [220], [267], [302]. Some publications derived summary metrics from their "expected trajectory" [86], [143], [220].

Fig. 4 illustrates how "expected trajectory" curves can be misleading. In this simple example, only the resistive duration is random. Integral-based metrics derived from the "expected trajectory" may reflect the expected value of metrics applied to constituent curves. However, magnitude-, duration-, rate-, and threshold-based metrics will not. Juul et al. highlights that summary metrics of expected trajectories can wildly differ from those of any single trajectory and suggests two alternatives: curve-based descriptive statistics and likelihoods of specific scenarios of interest [303].

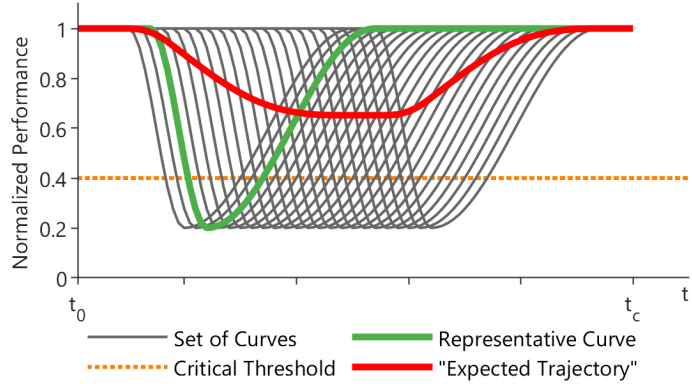

**Fig. 4.** *Example of how an "expected trajectory" curve provides misleading summary metrics*

Summary metrics may require special consideration when the scenario ensemble spans multiple hazard events. Hazard events may not be independent. Zhao et al. distinguished between "single-disruption" and "multiple-disruption" scenarios [65]. Zobel and Khansa provided metric adjustments for "partial resilience" [62]. In some cases, metric adjustment is not necessary. Integral-based metrics can extend their control time to encompass probabilistic hazards [31], [56], [69], [189], [220], [222], [230]. Alternatively, some authors enumerated distinct scenarios with probabilities, then weighted the aggregation of summary metrics [54], [65], [127].

## 4.8 Summary Functions

Within the survey, 21 publications implemented a function that does not map a resilience curve to a scalar value (i.e., the function is not a summary metric). Instead, these summary functions can be evaluated at any point within the scenario. Examples included *local resilience* as the derivative of the resilience curve [289], [290] and *space-time dynamic resilience measure* as the normalized cumulative performance since disruption [275]. These functions were often labeled a variation of "resilience" but such terminology is avoided in this manuscript.

**Recovery ratio**, illustrated in Fig. 3(b), quantifies the improvement of performance at any time relative to the curve's minimum performance. This was the most common summary function, with implementation across infrastructure sectors [19], [35], [38], [41], [210], [304]–[308]. Orwin and Wardle introduced it as *resilience index* [26], but Henry and Ramirez-Marquez established its most commonly used form: *value of resilience*, $Я(t)$ [68]. Recovery ratio was commonly implemented with a presumption of availability measures and a fixed $\mathcal{R}_0$ (e.g. restoration of maximum potential network flow [38], [41]), but Thekdi and Santos extended its definition to encompass performance targets, $\mathcal{R}(t)$ [94]. Since the function can be evaluated at any time, recovery ratio can imply that "resilience" is improving over time [38]. The metric provides $Я(t_d) = 0$ in all scenarios—this is not "zero resilience". Nor does $Я(t) = 1$ indicate the system is "fully resilient" [39], only that the system has recovered. Expressed with these considerations, recovery ratio may be useful for restoration sequencing [40], [41] and component importance estimation [36], [38], [278].

# 5 Discussion

Infrastructure resilience analyses commonly focus on prioritizing recovery actions, recommending system configurations or interventions, or supporting related analyses. Resilience curves are a useful tool to achieve such objectives; however, incorrectly using this tool can yield incorrect results. The following section discusses four aspects of resilience curves that should be carefully considered before their use: selection and implementation of performance measures, selection and implementation of summary metrics, communication of results, and improving the practice of resilience analysis.

## 5.1 Performance Measure Selection and Implementation

Each performance measure category can be loosely related to an analytical focus and applicability. While expanding the scope of an analysis can increases its complexity, it may open up additional opportunities for resilience-improving interventions.

- **Availability measures** focus on analysis on the infrastructure system itself. This is appropriate when the infrastructure's utilization is tightly coupled to its availability or when stakeholders are indifferent to infrastructure utilization. Such analyses will generally not need a model of utilization or service demand.
- **Productivity measures** incorporate both the supply and demand of infrastructure services. This level of analysis is expected for downstream stakeholders (i.e., customers). Analyses likely require a model of both the infrastructure system (i.e., supply) and its utilization (i.e., demand). This scope provides additional opportunities to improve system resilience, such as demand-response management.
- **Quality measures** expand supply and demand considerations into a representation of the service's character. Within these analyses, modeling effort may be dominated by non-infrastructure considerations (e.g., dynamics of service utilization). This provides a corresponding increase in intervention options. While not appropriate for all systems or scenarios, quality measures can be critical for comparing trade-offs between steady-state and disrupted performance. These measures may also be appropriate for smaller-scale disruptions in which the system is stressed such that quality, but not production, is reduced.

**Availability measures** focus an analysis on the infrastructure system itself and do not incorporate variation in the system's provision of service. This may be appropriate when stakeholders are solely interested in the system itself (e.g., utility operators, public works departments) or for scenarios in which full recovery is relatively quick (e.g., when stakeholders are not expected to change service demand or behavior). When availability measures are appropriate and stakeholders are focused on full restoration, a fixed reference, $\mathcal{R}_0$, and static normalization are reasonable. Modeling an infrastructure system under such assumptions is generally more straightforward than alternatives, but measures and assumptions should seek to reflect stakeholder goals and system realities—not modeling considerations. If inappropriately adopting availability measures and static normalization, analyses risk overvaluing excess capacity and undervaluing dynamic, real-time resilience capabilities.

Analyses should anticipate **productivity measures** as being of primary interest. Such measures parallel the definition of infrastructure: systems that "produce and distribute…essential goods and services" [183]. The ability to provide service is generally dependent on the system's availability, so productivity measures are often an extension of availability-based analyses. This extension is appropriate for stakeholders that are primarily concerned with the provision of service to users, such as public officials or the customers themselves. Additionally, productivity measures are only understood relative to demand on the system. While some systems are subject to constant service demand, most practical infrastructure supports varying loads (e.g., hourly, daily, and seasonal cycles in energy consumption).

Service demand should be assumed to be dynamic relative to disruption scenarios and system behavior, until demonstrated otherwise. This relates productivity measures to performance targets, $\mathcal{R}(t)$, and endogenous normalization. Starting with this expectation, exogenous, $\hat{\mathcal{R}}(t)$, or constant baselines, $\mathcal{R}_0$, should only be used when justified. This is in contrast to the existing literature, in which fixed references, $\mathcal{R}_0$, is often assumed without justification. The benefits of modeling simplicity cannot always be sufficient: when an analysis makes recommendations from $p$; the numerator, $\mathcal{P}(t)$, and denominator, $\mathcal{R}(t)$, are both relevant. While this increases the complexity of an analysis, it also introduces opportunities for resilience improvements. For example, a system may anticipate and prepare for disruptions, in response to a warning event, and reduce expected demand on an infrastructure, $\mathcal{R}(t)$, to minimize the normalized effect, $p(t)$, of reduced productivity, $\mathcal{P}(t)$. Such opportunities are broadly underexplored within quantitative, scenario-based resilience analyses.

**Quality measures** may provide the best representation of agent desires within the system (e.g., "how long is my travel time?" instead of "can I reach my destination?"). Such measures may shift analytical focus away from the infrastructure to supported systems, yet quality is still often dependent on both availability and productivity of the infrastructure. Quality measures may not be appropriate for all scenarios. If productivity goals cannot be met, then quality goals may not be relevant. From this perspective, quality measures may be appropriate for systems under less stress than those described with productivity measures. Quality measures may also be applicable when a disruption does not actually affect productivity measures. Performance targets for quality measures are likely to be constant, representing typical or desired levels. Unlike availability and productivity measures, quality measures may not meet their performance target during steady-state conditions (e.g., traffic delay is non-zero on typical workdays). This provides opportunities to assess resilience improvements as tradeoffs between steady-state and disrupted situations.

It is not always the case that availability measures demand less analytical effort than alternative measures. Consider an availability measure "households with water". Determining this measure may require modeling service demand—a productivity consideration—and water pressure—a potential quality measure. Ultimately, selection of the measures should be driven by stakeholder goals and capabilities, and analytical effort should follow.

**Ensemble measures** or indices may be appealing when faced with multiple candidate measures (e.g., the performance at multiple spatial-locations across an infrastructure system), but they are not without challenges. Ensemble measures may obfuscate nuances of constituent measures (e.g., if measures diverge in edges cases). When ensembles are used in software-based analyses, such nuances could be detected with the ensemble measure's partial derivates across simulated trajectories, which stakeholders could interpret as the marginal benefit for improving the constituent measure. This framing can highlight differences in ensemble measure formulations—specifically between addition and multiplication when a constituent measure nears zero. As an additional consideration, weighting between constituent measures may be endogenous, in a parallel to performance targets. Generally, these considerations are unexplored across the literature, but should be addressed by any analysis incorporating an ensemble performance measure.

## 5.2    Summary Metric Selection and Implementation

Summary metrics should be selected to best describe stakeholder goals. The taxonomy of metrics presented in this manuscript, with their considerations, seeks to aid analysts and stakeholder in defining metrics. Relevant to all categories are the analysis's control interval and curve-specific milestones. Integral-based metrics, in particular, must deliberately consider the interval upon which they are calculated. This interval should specifically avoid defining boundaries in terms of curve milestones (e.g., initial system disruption and system recovery) as those milestones could shift across a variety of curve trajectories.

Summary metrics definitions should consider the variety of resilience curve trajectories. Empirical or simulated curves may not match preconceived expectations (e.g., instantaneous performance loss, non-decreasing recovery). Metric calculation will often require criteria for curve milestones (e.g., the interval upon which to calculate failure rate). Some metrics may be undefined for some curves (e.g., recovery duration for unrecoverable systems). If metric definitions are not clearly articulated and validated, unexpected consequences may be hidden within a larger analytical effort—especially when metrics contribute to simulation and optimization. Within the survey, many publications defined metrics with an assumption of static normalization. Such assumptions, if established early within an analysis, may constrain the overall effort and its applicability. This may be avoided by clearly distinguishing between metrics derived from unnormalized $\mathcal{P}(t)$ and normalized $p(t)$.

Ensemble metrics face the same challenges as ensemble measures. The methods to determine and validate $f(J_1(p(t)), J_2(p(t)), ...)$ and $f(J(p_1(t)), J(p_2(t)), ...)$ are generally underexplored. Like ensemble measures and performance targets, weighting within such functions may be endogenous.

In general, time and performance thresholds are underexplored with regard to summary metrics. Not only can they provide stand-alone metrics, but they may influence the calculation of other metrics, indicating a non-linear translation in stakeholder value. However, incorporating thresholds introduces another consideration for modeling complexity—time and performance thresholds may be dynamic within an unfolding scenario. Like performance targets and weighting between measures or metrics, modeling endogenous thresholds requires additional focus on the dynamics between infrastructure, stakeholder behavior, and stakeholder goals.

## 5.3   Communication of Results

Just as analyses should reflect stakeholder goals, communication of results should consider stakeholder interpretation. For this reason, analysts may wish to avoid labeling performance measures and summary metrics as "resilience." Such terminology was particularly common for ensemble measures and cumulative performance metrics. Resilience cannot be described at a single point in time (i.e., a measure) or from a single resilience curve trajectory (i.e., a metric). Instead, measures and metrics quantifying resilience considerations; descriptive terminology can reflect that role.

When communicating results in tables or illustration, analysts may consider presenting both unnormalized and normalized results. In many cases, analysis was accomplished entirely with normalized values. This is reasonable within the paradigm of availability metrics and static normalization, but it is challenged with the adoption of performance targets and endogenous normalization. Changes in normalized values may reflect changes in the performance measure or its target. Interpretation of results may depend on this distinction. For quality measures, performance targets may represent ideal or typical performance. This slight distinction in "nominal" may be obfuscated when normalized values are presented alone. Finally, comparing multiple performance measures with their normalized values may minimize relative differences in their actual values (e.g., restoration of utilities in communities of vastly different size).

Summary metrics quantify a singular resilience curve. When considering an ensemble of scenarios (e.g., Monte Carlo simulation runs), results may best be described with the distribution of their summary metrics and descriptive statistics. Illustrating such distributions provides opportunities: consideration of overlap when comparing options and more intuitive understanding of variance and skew. Alternatively, despite its relatively frequent implementation, converting an ensemble of curves into an "expected trajectory" provides significant analytical risks. Such representations are potentially misleading to stakeholders, and metrics derived from the "expected trajectory" may be objectively incorrect.

## 5.4 Improving the Practice of Resilience Analysis

Infrastructure systems are interdependent, interact with agents and the environment, and operate outside "normal" bounds during the periods of interest. In modeling and simulation, there are very real challenges in balancing complexity, fidelity, tractability, and legibility. This survey has revealed common assumptions to be addressed when expanding resilience analysis efforts.

Performance targets, time and performance thresholds, and weighting between performance measures and metrics may all be endogenous. Understanding each, then representing them within an analysis, requires stakeholder engagement and validation. Modeling mechanisms may lie outside traditional infrastructure disciplines (e.g., linear time-invariant modeling). Candidate representations may require validation with empirical data on element-level interactions and system-wide behavior. In contrast, few of the publications in this survey provided empirical analysis; those that did focused on well-documented historical events.

Empirical analysis and calibration, more generally, warrants additional research. While "stakeholders" is used liberally in this manuscript, most systems have multiple stakeholders, for which goals and values may not fully align. Identifying stakeholders, in and of itself, is a worthy research effort. Each stakeholder may perceive the system through a specific set of measures and metrics. When these assessments contradict between stakeholders, an additional level of complexity is introduced. This topic is generally underexplored, likely due to assumptions of non-decreasing recovery and the common, yet unfounded, use of availability measures and static normalization. Under those assumptions, all resilience curves are monotonic transformations of one another, eliminating the chance of diverging results. If used to compare strategies, options are "less good" than one another, but never "bad". With deliberate scrutiny, these assumptions may not hold for real-world systems.

These research directions introduce ever-expanding complexity into infrastructure resilience analysis. However, they also provide new opportunities for improving each system's resilience. If endogenous performance targets are applicable in an analysis, then adjusting performance targets can be used to improve system resilience. If stakeholders' preferred performance measures diverge, then aligning their interests may support other resilience improvements. These are exciting opportunities for bettering real-world systems. Finally, this survey made clear that new models and metrics are not needed "for illustration purposes"—the field is ready for direct, validated, actionable analysis of practical systems and problems.

# 6 Conclusion

In reviewing 273 publications, this manuscript supports future critical infrastructure analysis by defining taxonomies for resilience curve performance measures and summary metrics. Recommendations for selecting measures and metrics based on the taxonomy are distilled from a critical review of the literature.

Three categories of performance measures are defined, in order of increasing modeling complexity: availability (capacity or aggregated function of the system), productivity (quantity of service provided by the system), and quality (character of service provided by the system). Spatial variations and/or multiple stakeholders may necessitate the use of multiple measures and their ensembles. Use of performance measures are further clarified by describing normalization schemes as static, exogenous, or endogenous. Static normalization is generally associated with availability measures, for which full system recovery is the goal. In contrast, productivity measures may demand dynamic performance targets, thus requiring models to relate infrastructure states to stakeholder goals and behaviors.

Summary metrics distill a curve to single value to facilitate comparison of multiple curves. This manuscript defines six categories of summary metrics: magnitude, duration, integral, rate, threshold, and ensembles. These metrics should be carefully selected and specified, as preconceived expectations common in the literature (e.g., instantaneous performance loss, non-decreasing recovery) can have a significant effect on the effectiveness of summary metrics to communicate system resilience to stakeholders. For an ensemble of curves, analyses should provide descriptive statistics on the distribution of each curve's metric—metrics should not be derived from an "expected trajectory" curve.

Throughout this manuscript, examples arise that illustrate that infrastructure systems are socio-technical systems. Existing literature focused on the technical aspects, leaving a need for future research into the social aspects and their interactions with the technical. One clear area where this is relevant is engaging stakeholders in the resilience analysis process, especially in the definition of "performance". Future research could validate analytical interpretations of stakeholder performance definitions, and the effect on resulting resilience recommendations. Resilience curves are just one tool to be used in resilience analysis. This manuscript aims to improve the design and use of this tool as a way to succinctly and effectively communicate resilience analysis methods and results to stakeholders.

# Acknowledgements

This research did not receive any specific grant from funding agencies in the public, commercial, or not-for-profit sectors. The authors thank their colleagues from the Energy Systems Group, Lincoln Laboratory and the Global Resilience Institute at Northeastern University for discussion, comments, and support to this work. The views expressed in this article are those of the authors and do not reflect the official policy or position of the United States Air Force, Department of Defense, or the U.S. Government.